\documentclass[twocolumn,floatfix, showpacs]{revtex4}

\usepackage[final]{epsfig}
\usepackage{amsmath}
\begin{document}
 
\title{A Theoretical Assessment of Jet-Hadron Correlations}
 
\author{Thorsten Renk}
\email{thorsten.i.renk@jyu.fi}
\affiliation{Department of Physics, P.O. Box 35, FI-40014 University of Jyv\"askyl\"a, Finland}
\affiliation{Helsinki Institute of Physics, P.O. Box 64, FI-00014 University of Helsinki, Finland}


\begin{abstract}
With high-statistics data coming from both RHIC and LHC, the experimentally available selection of hard tomographic probes for the medium created in ultrarelativistic heavy-ion (A-A) collisions is rapidly expanding. Jet-hadron (jet-h) correlation measurements as introduced by the STAR collaboration are a promising avenue to study the structure of highly modified jets in a differential way since the away side correlation can be measured down to very low transverse momenta ($P_T$) in an essentially hydrodynamical regime. At the same time the geometry bias introduced by the trigger condition ensures that the away side shower has propagated a long distance in the medium. The aim of this paper is to provide a theoretical overview of the observable, discuss similarities and differences to other correlation observables such as hadron-hadron (h-h) and gamma-hadron ($\gamma$-h) correlations and to provide recommendations for future measurements.
\end{abstract}
 
\maketitle

\section{Introduction}

While 'jet quenching' has been the label for the physics of hard perturbative Quantum-Chromodynamics (pQCD) probes in the context of ultrarelativistic heavy-ion (A-A) collisions for years, for a long time many observables such as the single inclusive hadron spectrum and its nuclear modification factor $R_{AA}$  were really only sensitive to the fate of the leading shower parton in a jet. However, fully reconstructed jets in an A-A environment were among the first high $P_T$ observables at the LHC in terms of the so-called dijet imbalance $A_J$ \cite{ATLAS,CMS}, marking a pronounced transition point in the field from a focus on leading hadron observables to jet observables.

Yet, the first detailed characterization of the complete medium-modified jet structure in A-A collisions has been performed by the STAR collaboration at RHIC using triggered jet-hadron correlations \cite{STAR-jet-h}. This measurement can be classified among other triggered correlation measurements at RHIC, for instance h-h correlations \cite{STAR-DzT} or $\gamma$-h correlations \cite{gamma-h-STAR,gamma-h-PHENIX} in which the trigger condition provides information about the hard process whereas the medium modification of the shower is inferred from the recoiling (away side) parton remnants.

Mathematically, triggered correlations always involve a conditional probability of some process given a near (trigger) side fulfilling the trigger conditions. For this reason, their results are not straightforward to interpret and sometimes even counter-intuitive. However, triggered correlations are powerful measurements probing many different aspects of the hard process and its final state interaction with the medium. It is the aim of this work to provide the theoretical background for the interpretation of jet-h correlations and contrast their capabilities with triggered h-h and $\gamma$-h correlations.

The paper is organized as follows: After a short introduction to the observable and its terminology, a general overview of the possible biases introduced by the trigger condition is given. Since all comparison of theory results with the data is done using the in-medium shower evolution Monte-Carlo (MC) code YaJEM \cite{YaJEM1,YaJEM2,YaJEM-D} in its latest version YaJEM-DE \cite{YaJEM-DE}, a short summary of the model is provided before discussing the physics of near and away side in triggered jet-h correlation in comparison with h-h correlations. Finally, a comparison with the STAR data is made before drawing conclusions about the relevant physics and possible future refinements of the measurement. 

\section{Overview of high $P_T$ correlation observables}

\subsection{The physics of hard correlations}

In leading order (LO) pQCD, a hard scattering process results in a highly virtual back-to-back parton pair. These 'parent' partons develop into showers of daughter partons with progressively decreasing virtuality scale until a non-perturbative scale is reached, at which point non-perturbative dynamics changes the parton showers into jets of hadrons. The total momentum balance of these two jets contains information about the primary hard QCD process whereas the momentum distribution of hadrons inside each the jets contains information about the QCD dynamics of partonic shower, hadronization and any medium modification. Based on uncertainty relation arguments, it is expected that the medium modifies the evolution of a parton shower and the hadronization process at low $P_T$, but not the hard process itself or hadronization of the leading hadrons which due to time dilatation have a long formation time exceeding the medium lifetime. In particular, a systematic analysis of high $P_T$ observables \cite{Constraining} suggests that energy flows from the leading shower partons into the production of a broad and soft tail of subleading hadron production and to a small fraction of about 10\% also into direct excitation of medium degrees of freedom.

Imposing a trigger condition involving a sufficently large momentum scale picks the remnants of one parton out of the pair, ensures that a hard process has taken place and constrains the kinematics of the original parton pair to some degree. However, no trigger condition selects jets in an unbiased way, and the requirement that a shower fulfills the trigger condition selects a subset of shower evolutions out of all possible evolutions with properties determined by the precise details of the trigger condition. The bias may affect the parton momentum distribution, as well as a particular geometrical configuration inside the medium or correspond to a selection effect on parton type (see below).

The strength of the trigger bias in the absence of a medium is difficult to assess in a model-independent way, but the additional bias introduced by the medium modification of showers correlates directly with the single hadron/jet suppression factors $R_{AA}$: choosing a smaller subset out of the available set of all shower evolutions implies a stronger bias and at the same time a suppression of the rate of triggered events relative to the p-p case. Measuring the medium-induced suppression factor of the trigger rate is thus an efficient way of monitoring the total medium-induced bias. 

The distribution of correlated hadrons on the away side opposite to the trigger allows thus to study the modification of a biased subset of showers given the trigger conditions, but without additional biases introduced on the away side.

\subsection{Observables and terminology}

The STAR experiment clusters particles into jets using the following conditions: 1) only $\pi^+, \pi^-, \pi^0, K^+, K^-, p, \overline{p}$ and $\gamma$ can contribute to jets \cite{Joern} 2) all particles (i.e. tracks or calorimeter towers) are required to have $P_T > 2$ GeV 3) due to the use of a high tower trigger for the events, the trigger jet is required to have at least one tower with $P_T > 6$ GeV. 

The resulting particles are clustered into jets with the anti-$k_T$ algorithm with a radius parameter of $R=0.4$ using the FastJet package \cite{FastJet}. 
Background fluctuations are accounted for by reweighting the p-p jet distribution to match the equivalent Au-Au jet $P_T$ distribution.  The reweighting factors are determined by embedding p-p jets into Au-Au events \cite{Alice}. If the leading jet of the event falls into a certain momentum range, the trigger range (in practice, 10-15 GeV and 20-40 GeV are used), the trigger condition is fulfilled.

The direction of the trigger jet defines the near side hemisphere ($\phi = 0$), the hemisphere opposite to the trigger jet in azimuth ($\phi \approx \pi$) where the correlated hard parton remnant is expected is referred to as the away side hemisphere. In principle, the away side jet can be in a large rapidity range, in practice kinematics is such that both near and away side jets are distributed in narrow peaks around midrapidity (the rapidity of the away side jet is however an issue for current jet measurements at LHC where the distributions are significantly wider).

The primary observables are the yield per trigger (YPT) of hadrons on the away side and the Gaussian width of the away side correlation signal around $\phi = \pi$ as a function of away side (associate) momentum. From the YPT, the conditional away side suppression factor 
\begin{equation}
I_{AA}(P_T) = YPT_{AA}(P_T)/YPT_{pp}(P_T)
\end{equation}

and the momentum balance function
\begin{equation}
D_{AA}(P_T) = YPT_{AA}(P_T) \langle P_T \rangle_{AA} - YPT_{pp}(P_T) \langle P_T\rangle_{PP}
\end{equation}

can be derived. Note that absolute normalization uncertainties in the conditional yields cancel in $I_{AA}$ whereas they persist in $D_{AA}$. 

Since any background not correlated with the trigger jet creates a signal which is uniform in $\phi$ when averaged over many events, a triggered correlation allows to follow the fate of the away side jet to very low $P_T$ hadrons and to large angles without running into the problem of separating jet and background explicitly which is present in observables based on analyzing fully reconstructed jets\cite{A_J}.

As a side remark, this does not imply that the medium background can be neglected: Hard processes are for instance correlated with the reaction plane since the average in-medium pathlength is smaller in-plane than out of plane, which leads to a stronger relative suppression of trigger rates out of plane. But at the same time, the bulk medium momentum distribution is correlated with the reaction plane because the pressure gradients driving its fluid-dynamical evolution are stronger in-plane than out of plane. Thus, since both jets and bulk are correlated with the reaction plane, the background medium is in fact correlated with the jet (although this is not a causal correlation, the jet does not cause the bulk medium to to align with its direction), and hence a flow modulation of the background medium needs to be subtracted from the correlated yield to find the correlations which are caused by the jet and represent its energy redistribution.

\subsection{Biases}

Let us give a short overview in what way the requirement of a trigger hadron may bias the away side (an expanded version of the discussion, which is general for any triggered correlation, can be found in \cite{h-h}).

In vacuum, the relevant effects are kinematic bias and parton type bias. They are caused by the fact that usually a fraction of the energy carried by the full  jet will not fall into the experimental definitions by particles being either too soft to fall above the $P_T$ cut or being at an angle larger than $R$ and thus falling outside the cone. For given parton energy, the highest energy fraction will be recovered if the fragmentation pattern is hard and collimated. For a parton energy distribution according to the pQCD production spectrum, this translates into a systematic average offset $\Delta \epsilon$ between experimental jet energy and underlying parton energy determined by the competition between the probability of having a hard fragmentation of a frequently available relatively low mometum parton or a soft fragmentation of a rare high energy parton. Since this competition is driven by the steepness of the pQCD parton spectrum, the bias is very different between RHIC and LHC.

Another aspect of the kinematic bias is that any momentum imbalance on the partonic level (usually referred to as 'intrinsic $k_T$' and taken to account for initial state as well as higher order pQCD effects) is likely to be oriented into the trigger direction, leading on average to a difference between near side and away side parton energies. 

The parton type bias is related: Since quark jets tend to be harder and more collimated than gluon jets, everything else being equal the trigger condition is more likely fulfilled for a quark jet than for a gluon jet (of course, this has to be discussed in the context of the \emph{a priori} probability to produce a hard gluon vs. a hard quark, which in the low $P_T < 40$ GeV range at LHC still implies that a significant number of triggers would be gluons). Since the dominant hard channel in the RHIC kinematic range is $qg \rightarrow qg$, the parton type bias to trigger on a quark translates into a bias for the away side parton to be a gluon (again, this is different for the LHC low $P_T$ range where the dominant channel is $gg\rightarrow gg$).

In the presence of a medium, generically jets are broadened and softened proportional to the length of their path through the medium and the interaction strength between parton and medium. This has several implications:

The kinematic bias is medium-modified since the fraction of energy falling into the experimental jet definition is decreased by medium-induced broadening and softening of jets, i.e. $\Delta \epsilon$ is increased in the medium. This implies that the away side parton in the medium has generically \emph{larger} energy than in vacuum, which tends to lead to a counter-intuitive increase of the away side yields as compared to the vacuum.

The kinematic bias now correlates with a geometrical bias to have a short in-medium pathlength (and hence only little modification) for the trigger. This geometrical bias distorts the {\itshape a priori} distribution of hard vertices (which follows binary collision scaling) to a more surface-biased distribution. This in turn translates into a larger-than-average in-medium pathlength for the away side parton, i.e. stronger medium modifications. 

The medium also strengthens the parton type bias, since a gluon interacts by a factor of 4/9 more strongly with the surrounding color charge. This means that in medium a quark is even more likely to lead to a trigger as compared with a gluon and increases the probability to find a quark on the near and a gluon on the away side even beyond the bias in vacuum. This is sometimes referred to as 'gluon filtering'.

Both geometrical and parton type bias in medium tend to lead to a suppression of away side yields. The actual away side yield modification depends on a non-trivial cancellation of kinematic, geometrical and parton type bias, which in turn depends on the kinematics, relevant pQCD channels and strength and geometry of the medium. It is this dependence on multiple quantities of interest which makes correlation observables powerful and interesting.

\subsection{Qualitative differences of jet-h to triggered h-h and $\gamma$-h correlations}

At this point, we may try to establish some differences between the triggered objects in correlation measurements. In all jet-h, h-h and $\gamma$-h correlations, the trigger represents a proxy of the near side parton from which the away side kinematics is inferred, but it does so in different ways.

A $\gamma$ has the closest relationship between triggered object and away side parton energy --- in LO, up to intrinsic $k_T$ imbalance, the photon energy equals the away side parton energy. However, NLO effects and the creation of photons in the parton shower will in general dilute the connection somewhat. A $\gamma$-h correlation is thus characterized by a very small $\Delta \epsilon$ and the parton type bias dictates that the overwhelming number of away side partons are quarks since $gq \rightarrow \gamma q$ is the dominant reaction channel. Since the photon does not interact strongly, there is in principle no medium modification to the kinematic and parton type bias and no geometrical bias at all. However, again the presence of fragmentation and jet conversion photons can potentially change this.

Since any hard single hadron corresponds to a hard jet, but not every jet contains a hard hadron, the subclass of shower evolutions leading to a triggered jet is significantly larger than the subclass leading to a triggered hadron in the same energy range. In general, the fact that subleading hadrons are clustered into a jet means that the kinematic bias in vacuum as well as the parton type bias is weaker for jets than for hadrons. 

When embedded into a medium, this may or may not be true. Under LHC kinematical conditions with jet energies above 100 GeV, jets turn out to be remarkably robust against medium modifications and acquire for instance no significant geometrical bias \cite{A_J}. On the other hand, the RHIC condition of clustering only particles with $P_T > 2$ GeV implies for 10-15 GeV jets that each hadron needs to carry as much as about 20\% of the total jet energy. In vacuum this selects  events in which the jet energy is shared across several hard partons rather than a single hard and many soft partons as in the case of a hadron trigger. Such a configuration in the medium however can naively be thought of as multiple partons undergoing energy loss rather than a single parton, thus amplifying any medium modification. As a result, such jets can under some conditions be more suppressed and can acquire a stronger surface bias than single hadrons, a scenario which will be explored in detail in section \ref{SurfaceBias}.

\section{Theoretical modelling}

The theoretical modelling of jet-h correlations involves several building blocks: 1) simulation of the hard process 2) embedding of the evolving parton showers into a hydrodynamical medium and computing the medium modification to the shower evolution 3) clustering of the resulting hadron distributions into jets, including an approximate simulation of the background medium fluctuations 4) after evaluating the trigger condition, computation of the away side correlation yields and Gaussian width.

\subsection{The hard process}

In LO pQCD, the production of two hard partons $k,l$ 
is described by
\begin{equation}
\label{E-2Parton}
  \frac{d\sigma^{AB\rightarrow kl +X}}{dp_T^2 dy_1 dy_2} \negthickspace 
  = \sum_{ij} x_1 f_{i/A}(x_1, Q^2) x_2 f_{j/B} (x_2,Q^2) 
    \frac{d\hat{\sigma}^{ij\rightarrow kl}}{d\hat{t}}
\end{equation}
where $A$ and $B$ stand for the colliding objects (protons or nuclei) and 
$y_{1(2)}$ is the rapidity of parton $k(l)$. The distribution function of 
a parton type $i$ in $A$ at a momentum fraction $x_1$ and a factorization 
scale $Q \sim p_T$ is $f_{i/A}(x_1, Q^2)$. The distribution functions are 
different for free protons \cite{CTEQ1,CTEQ2} and nucleons in nuclei 
\cite{NPDF,EKS98,EPS09}. The fractional momenta of the colliding partons $i$, 
$j$ are given by $ x_{1,2} = \frac{p_T}{\sqrt{s}} \left(\exp[\pm y_1] 
+ \exp[\pm y_2] \right)$.
Expressions for the pQCD subprocesses $\frac{d\hat{\sigma}^{ij\rightarrow kl}}{d\hat{t}}(\hat{s}, 
\hat{t},\hat{u})$ as a function of the parton Mandelstam variables $\hat{s}, \hat{t}$ and $\hat{u}$ 
can be found e.g. in \cite{pQCD-Xsec}.

To account for various effects, including higher order pQCD radiation, transverse motion of partons in the nucleon (nuclear) wave function, the distribution is commonly folded with an intrinsic transverse momentum $k_T$ with a Gaussian distribution, thus creating a momentum imbalance between the two partons as ${\bf p_{T_1}} + {\bf p_{T_2}} = {\bf k_T}$. 

We evaluate Eq.~(\ref{E-2Parton}) at midrapidity $y_1 = y_2 = 0$ and sample this expression using a MC code introduced in \cite{Dihadron} by first generating the momentum scale of the pair and then the (momentum-dependent) identity of the partons. A randomly chosen $k_T$ with a Gaussian distribution of width 2.0 GeV is then added to the pair momentum. This value is obtained in a best fit to the $YPT_{pp}(P_T)$ as measured by STAR \cite{Alice}.

\subsection{Embedding into hydrodynamics}

We assume that the distribution of vertices follows binary collision scaling as appropriate for a LO pQCD calculation. Thus, the probability density to find a vertex in the transverse plane is

\begin{equation}
\label{E-Profile}
P(x_0,y_0) = \frac{T_{A}({\bf r_0 + b/2}) T_A(\bf r_0 - b/2)}{T_{AA}({\bf b})},
\end{equation}
where the thickness function is given in terms of Woods-Saxon distributions of the the nuclear density
$\rho_{A}({\bf r},z)$ as $T_{A}({\bf r})=\int dz \rho_{A}({\bf r},z)$ and $T_{AA}({\bf b})$ is the standard nuclear overlap function $T_{AA}({\bf b}) = \int d^2 {\bf s}\, T_A({\bf s}) T_A({\bf s}-{\bf b})$ for impact parameter ${\bf b}$. We place the parton pair at a probabilistically sampled vertex $(x_0,y_0)$ sampled from this distribution with a random orientation $\phi$ with respect to the reaction plane.

The medium itself is described using an ideal 2+1d hydrodynamical model \cite{hydro2d} from which the energy density $\epsilon({\bf r}, \tau)$ is determined at each space-time point with transverse coordinate ${\bf r}$ and proper time $\tau$ (where space-time rapidity $\eta_s = 0$). Since it is  known that there is a sizeable dependence of several high $P_T$ observables on the choice of the underlying hydrodynamical model (most notably when the dependence on the angle with the reaction plane is concerned) \cite{JetHydSys}, we have verified that embedding into a 3+1d ideal model \cite{hydro3d} (which in \cite{JetHydSys} exhibited the strongest difference to \cite{hydro2d}) does not lead to substantially different results for the observables discussed in this work.

\subsection{In-medium shower evolution}

The evolution of parton showers in the medium is computed with the MC code YaJEM, which is based on the PYSHOW code \cite{PYSHOW} which in turn is part of PYTHIA \cite{PYTHIA}. It simulates the evolution from a highly virtual initial parton to a shower of partons at lower virtuality in the presence of a medium. A detailed description of the model can be found in \cite{YaJEM1,YaJEM2,YaJEM-D}. Here we use the version YaJEM-DE \cite{YaJEM-DE} which is one of the best-tested theoretical models available for in-medium shower evolution and gives a fair account of a large number of high $P_T$ observables both at RHIC and LHC \cite{Constraining,A_J,RAA-LHC,A_J_edep}. 

In YaJEM-DE, the medium is characterized by two transport coefficients, $\hat{q}$ and $\hat{e}$. Here, $\hat{q}$ parametrizes the virtuality growth of a parton per unit pathlength and leads to medium-induced radiation whereas $\hat{e}$ describes the energy loss of propagating partons into non-perturbative medium modes. These transport coefficients are assumed to be related to the energy density $\epsilon$ of the hydrodynamical medium as

\begin{equation}
\label{E-qhat}
\hat{q}[\hat{e}](\zeta) = K[K_D] \cdot 2 \cdot [\epsilon(\zeta)]^{3/4} (\cosh \rho(\zeta) - \sinh \rho(\zeta) \cos\psi)
\end{equation}

with $\zeta$ the position along the path of the propagating parton, $\rho$ the medium transverse flow rapidity, $\psi$ the angle between parton direction and medium flow vector and $K,K_D$ two parameters regulating the strength of medium-induced radiation vs. direct energy loss into the medium.
As in \cite{YaJEM-DE}, the free parameters $K, K_D$ are adjusted such that the energy loss into non-perturbative modes is a 10\% contribution to the total as constrained by a number of other observables \cite{Constraining}.

Given $(\hat{q}(\zeta),\hat{e}(\zeta))$ along the parton path, YaJEM computes the in-medium partonic shower evolution and hadronizes the result using the Lund string model \cite{Lund} so that the fragmentation remnants of the initial hard parton can be analyzed on the hadron level.

For the p-p baseline in the absence of a medium, YaJEM by definition reproduces the results of PYSHOW followed by Lund hadronization, i.e. it reduces to standard PYTHIA results.

\subsection{Clustering into jets}

The resulting hadrons which pass the PID cut and have $P_T > 2$ GeV are now clustered into jets if a hadron above 6 GeV is found in the event. Clustering is done using the anti-$k_T$ algorithm of the FastJet package \cite{FastJet} with $R=0.4$. The leading jet in the event is taken to be a trigger candidate. In order to account for medium background fluctuations, a random background energy term with a Gaussian distribution of 1 GeV width is added to the trigger candidate energy (note that due to the $P_T > 2$ GeV cut, background fluctuations are much suppressed).

Technically, it is easiest to cast the results of this procedure into the probability of recovering the energy $E_{jet}$ within the jet definitions given a parton with initial energy $E_0$ and the path $\zeta(\tau)$ of the parton through the medium. Using a scaling law established in \cite{YaJEM2}, the path can be replaced by the total line integrated virtuality $\Delta Q^2_{tot}$ along the path (since $\hat{q} \sim K_D/K \hat{e}$ the direct energy loss into the medium is implicitly covered by this procedure). The probability $P(E_{jet}|E_0,\Delta Q^2_{tot})$ can then be conveniently convoluted with the pQCD parton spectrum, allowing for a numerically fast evaluation of the trigger condition. 

If the trigger condition is fulfilled, the away side shower is computed on the hadron level and the $YPT(P_T)$ as well as the Gaussian width of the correlation signal are evaluated.

\section{Near side results}

In this section, we discuss some interesting and perhaps unexpected aspects of the biases induced by the trigger condition.

\subsection{Geometry bias}

\label{SurfaceBias}

\begin{figure*}[!htb]
\begin{center}
\epsfig{file=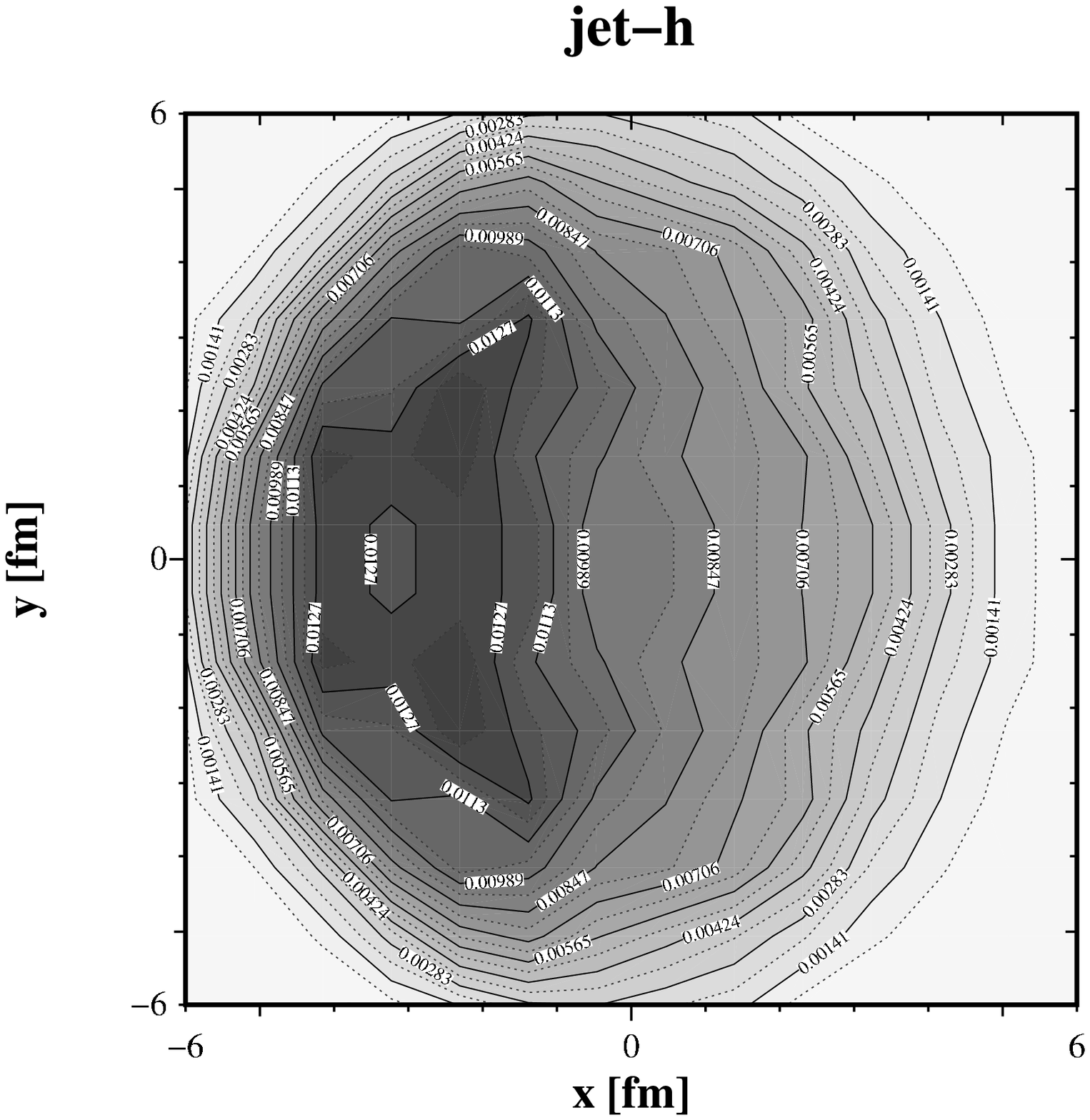, width=7cm}\epsfig{file=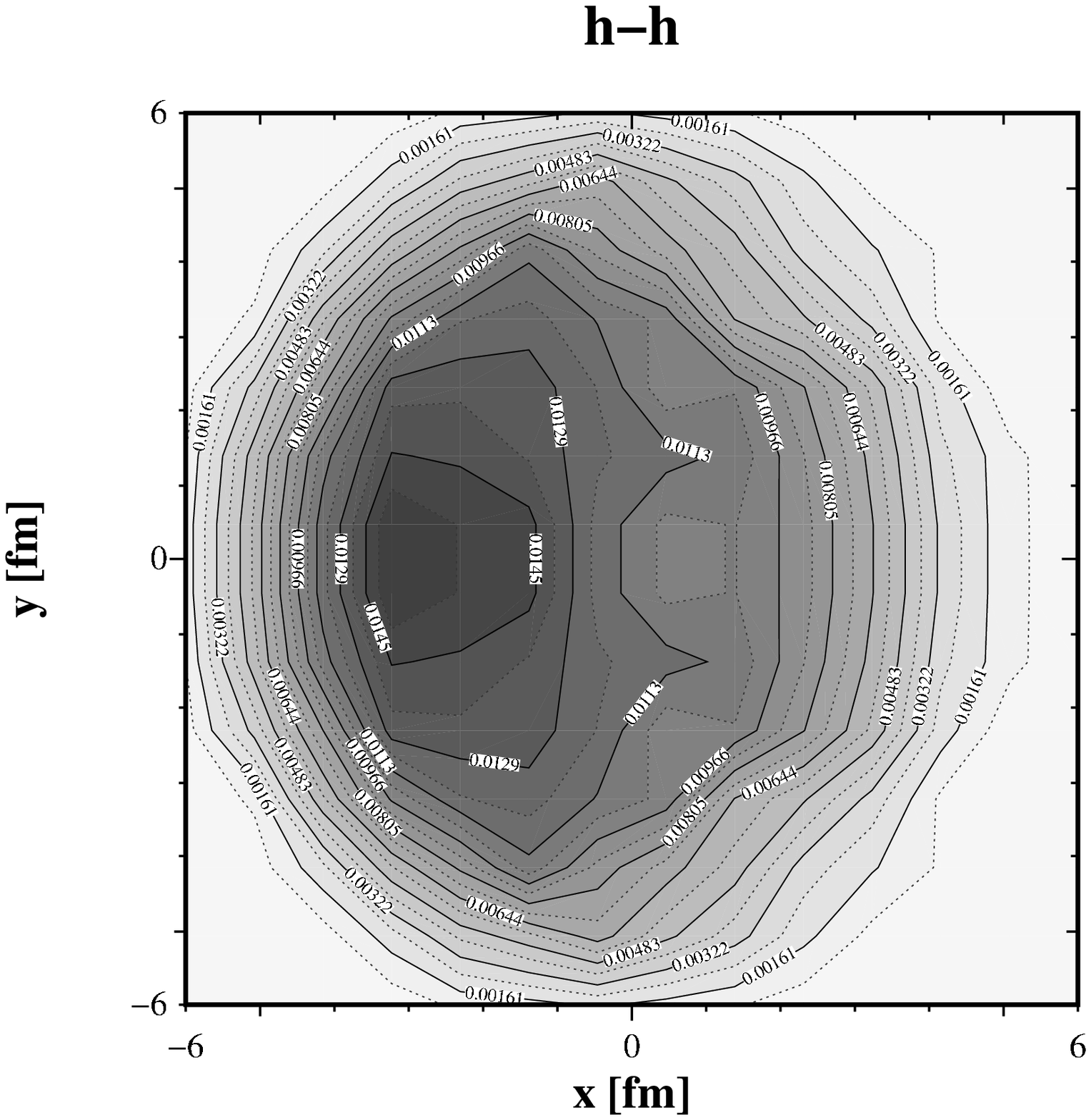, width=7cm}
\end{center}
\caption{\label{F-Geobias}Comparison of the probability density of a vertex in the transverse $(x,y)$ plane to fulfill a 10-15 GeV trigger condition in 0-10\% central 200 AGeV Au-Au collisions. Left panel for a jet trigger as used by STAR (see text), right panel for comparison for a single charged hadron trigger). In all cases, the trigger parton moves into the $-x$ direction.}
\end{figure*}

In Fig.~\ref{F-Geobias} the conditional probability distribution of finding a hard vertex given a triggered object in the 10-15 GeV range is shown for both a jet and a single charged hadron trigger. Somewhat surprisingly, the jet trigger is \emph{more} surface biased than the hadron trigger.
The visual impression can be quantified by introducing the variable $s = N_{near}/N_{away}$ where $N_{near}$ is the number of vertices found in the near side ($-x$) hemisphere whereas $N_{away}$ is the number in the away side hemisphere. For the jet trigger we find $s=2.08$ whereas for the hadron trigger we obtain $s=1.84$. 

This appears to be in manifest contradiction to the results of \cite{A_J} where a jet trigger for $R=0.4$ was found to be sigificantly \emph{less} surface biased than a single hadron trigger. However, it is crucial to take note of the differences in $P_T$-cut and jet energy. In \cite{A_J}, LHC jets measured by ATLAS with $E_{jet} > 100$ GeV were investigated while $P_T > 1$ GeV was required for each particle to be clustered, i.e. a single particle was required to carry about 1\% of the jet energy. For STAR, jet energy and cut are 10 GeV and 2 GeV respectively, i.e. a single particle is required to carry as much as 20\% of the jet energy. While the former condition is not restrictive, the latter favours event topologies with multiple hard partons, which in the medium undergo $n$ times the energy loss of a single parton. As a result, such jets are sensitive to medium modification beyond the leading hadron. An equivalent situation at LHC would require to cluster only particles with $P_T > 20$ GeV into jets --- such a cut is likely to create a highly geometry-biased sample. 

For the higher trigger range of 20-40 GeV, we find $s= 2.16$ for the jet trigger and $s=2.47$ for the hadron trigger, i.e. the situation reverses and the hadron trigger becomes more surface biased. This confirms the idea that the 2 GeV track cut becomes increasingly unimportant as jet energy increases.

\subsection{The $P_T$ cut illustrated}

In order to get a better insight into the effect of the constituent $P_T$ cut on jets, let us study the simple case of a fragmenting 20 GeV quark in vacuum and for a fixed path in-medium with $\Delta Q^2_{tot} = 5$ GeV$^2$. It is important to realize that imposing a $P_T$ cut suppresses both the vacuum and the in-medium rate of jets into a given energy range. However, the relative rate of jets in medium to vacuum (jet $R_{AA}$) is only affected by the cut if there is medium suppression due to the cut beyond what is already observed in vacuum (this is different from the suppression of the single inclusive hadron spectrum where there are no cuts which affect the vacuum rate). In addition, there is always the angular cut which suppresses both jets in vacuum and in medium as compared to the parton production rate unbiased by jet finding.

\begin{figure}[!htb]
\begin{center}
\epsfig{file=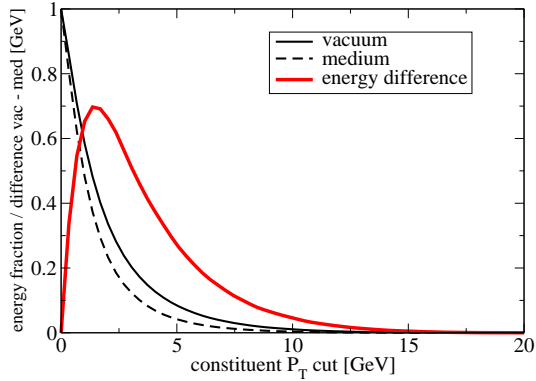, width=7cm}
\end{center}
\caption{\label{F-Cutbias} Relative fraction of the total jet energy in a cone of $R=0.4$ recovered as a function of constituent $P_T$ cut for vacuum (black solid) and medium-modified jet (black dashed) as well as energy difference between vacuum and medium jet induced by the cut (red solid) for a 20 GeV quark (see text).}
\end{figure} 

In Fig.~\ref{F-Cutbias} the fraction of energy within a cone of $R=0.4$ coming from a 20 GeV quark jet with the STAR PID cuts applied is shown as a function of the constituent $P_T$ cut. From the figure, it is evident that a large fraction of the jet energy for this kinematics is carried by hadrons below 3 GeV even in vacuum, and that the distribution is even softer in a medium-modified jet.

To study the effect of the $P_T$ cut on the jet rate suppression in medium, the energy difference between vacuum and medium case (i.e. the medium-induced energy radiated out of the jet definition) as a function of the $P_T$ cut is shown where the energy of the in-medium jet has been artificially normalized to the vacuum case at $P_T = 0$ to eliminate the effect of the cone radius cut.

It is evident that the difference peaks at about 1.5 GeV, i.e. applying a constituent cut of about 1.5 GeV makes a jet maximally sensitive to the additional softening of the fragmentation pattern in the medium and leads to the most significant medium-induced suppression. For a higher $P_T$ cut, both vacuum and medium case are very much suppressed, but there is little additional medium suppression. It is the fact that the 2 GeV cut applied by STAR is very close to the optimal 1.5 GeV which makes the resulting jet rate very sensitive to the effect of the medium.

\subsection{Kinematic bias}

\begin{figure*}[htb]
\begin{center}
\epsfig{file=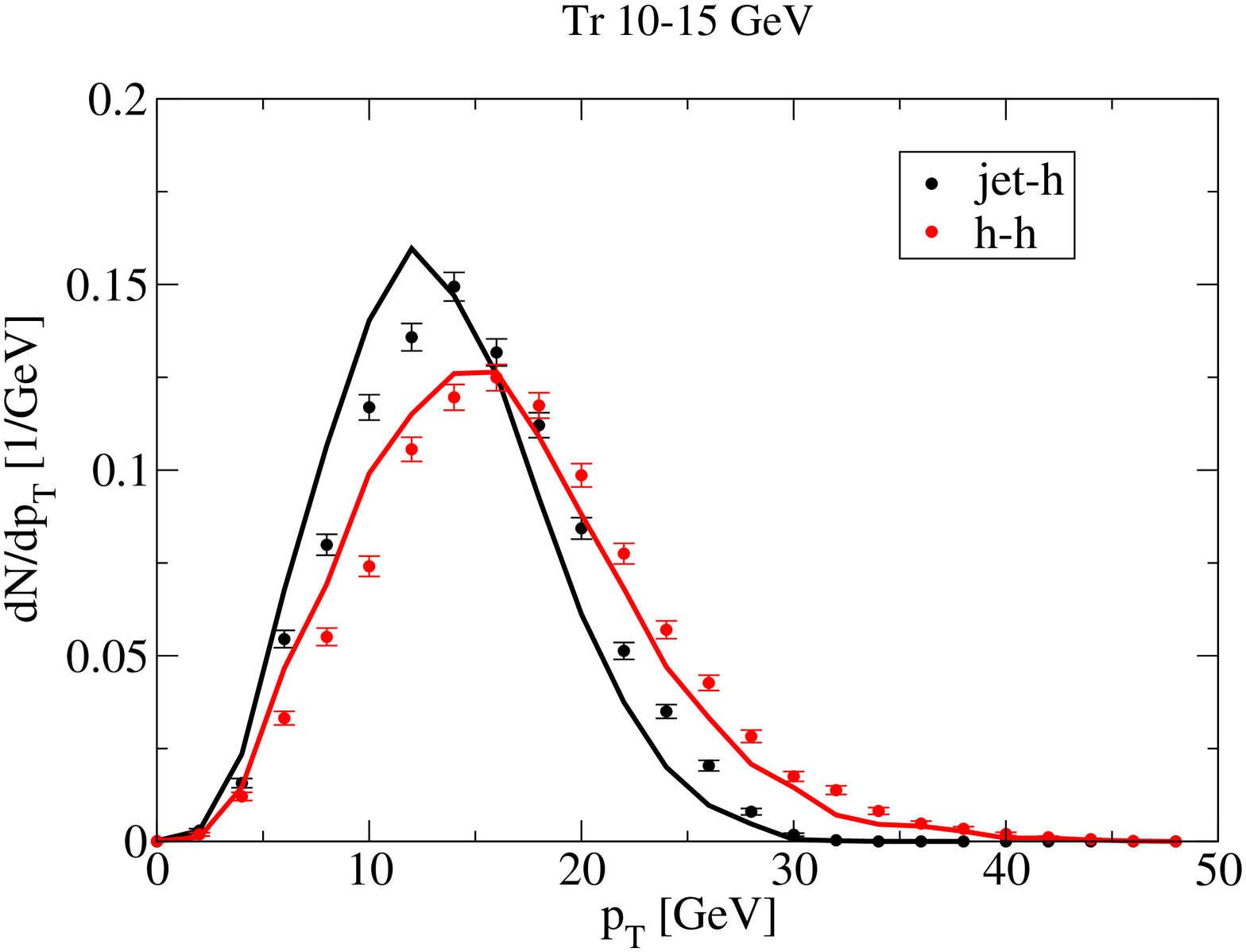, width=7cm}\epsfig{file=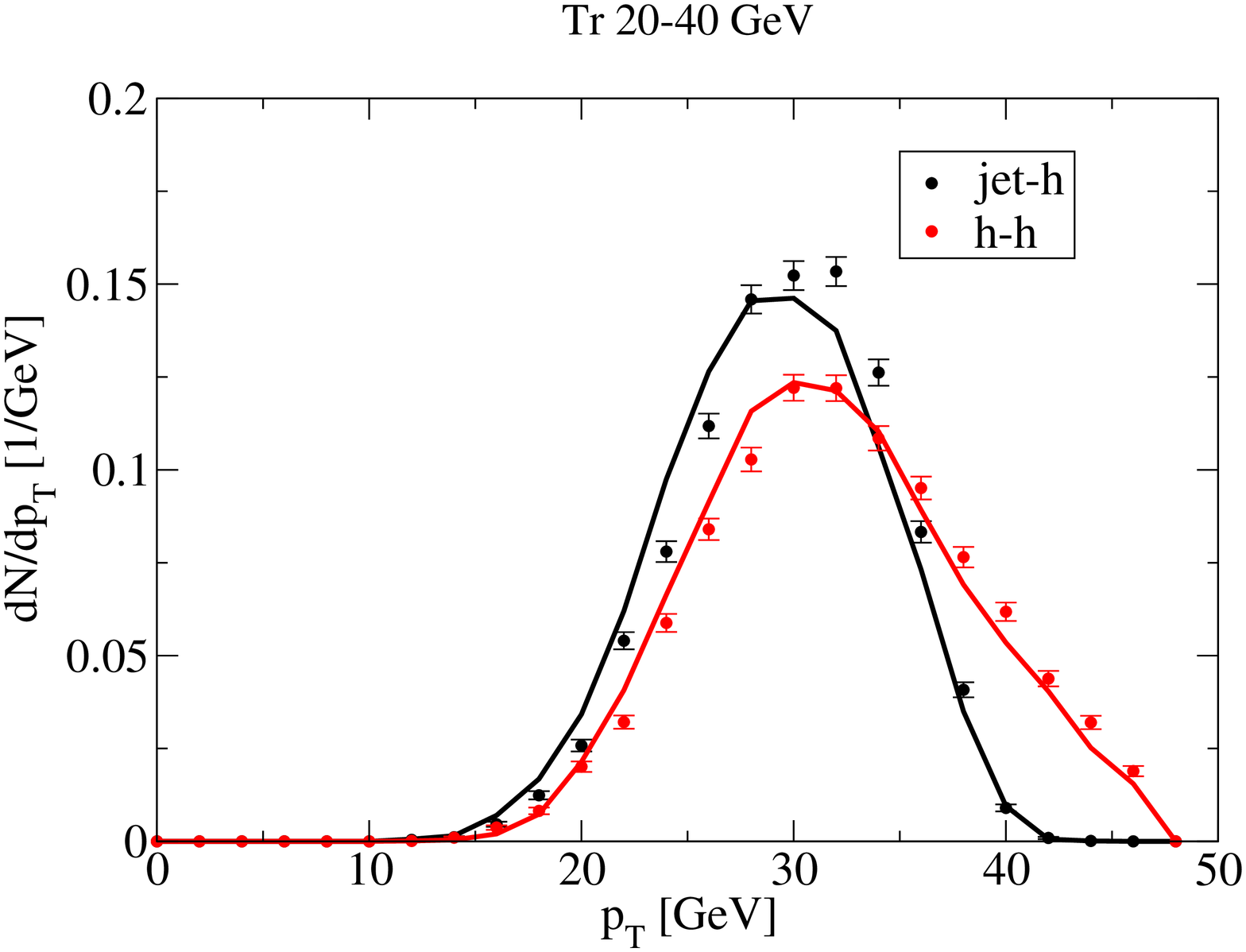, width=7cm}
\end{center}
\caption{\label{F-Kinbias}Conditional distribution of away side parton momenta given a triggered object in the 10-15 GeV range (left) and the 20-40 GeV range (right) for vacuum (lines) and medium-modified jets (points).}
\end{figure*}

In order to discuss the kinematic bias, it is useful to study the distribution of away side parton momenta given a triggered object. In the absence of higher order QCD effects, intrinsic $k_T$, shower evolution and background fluctuations in jet finding, the back-to-back partons are expected to have the same energy, i.e. the distribution should be a delta function at the trigger energy for vanishing trigger momentum bin width and smeared across the trigger range with a weight given by the parton production cross section as a function of momentum for any realistic situation. As can be seen from Fig.~\ref{F-Kinbias}, the actual distribution when all these effects are taken into account is a fairly broad Gaussian.

The probability of having a gluon jet on the near or away side $P_{glue}^{near}, P_{glue}^{away}$ along with the average momenum on near and away side and the Gaussian width of the away side momentum distribution as extracted from Fig.~\ref{F-Kinbias} is shown in Tables~\ref{T-Kinbias1} for a 10-15 GeV trigger range and \ref{T-Kinbias2} for a 20-40 GeV trigger range.

\begin{table*}
\begin{tabular}{|l|c|c|c|c|}
\hline 
{\bf 10-15 GeV trigger}& jet-h vacuum & jet-h medium & h-h vacuum & h-h medium\\
\hline \hline
$\langle P_T \rangle_{near}$ [GeV]  & 18.2 & 19.2 & 20.4 & 21.5\\
$P^{glue}_{near}$ & 0.58 & 0.29 & 0.05 & 0.05\\
$\langle P_T \rangle_{away}$ [GeV] & 14.3 &  15.5 & 16.5 & 17.5\\
Gaussian $P_T$ width [GeV] & 7.1 & 7.7 & 8.8 & 9.0\\
$P^{glue}_{away}$ & 0.48 & 0.60 & 0.69 & 0.69\\
\hline
\end{tabular}
\caption{\label{T-Kinbias1} Parameters characterizing the kinematic and parton type bias for a 10-15 GeV trigger range for hadron and jet triggers (see text).}
\end{table*}

\begin{table*}
\begin{tabular}{|l|c|c|c|c|}
\hline 
{\bf 20-40 GeV trigger}& jet-h vacuum & jet-h medium & h-h vacuum & h-h medium\\
\hline \hline
$\langle P_T \rangle_{near}$ [GeV]  & 32.9 & 33.7 & 35.4 & 36.0\\
$P^{glue}_{near}$ & 0.39 & 0.18 & 0.02 & 0.02\\
$\langle P_T \rangle_{away}$ [GeV] & 30.0 &  30.6 & 32.8 & 33.6\\
Gaussian $P_T$ width [GeV] & 7.5 & 7.1 & 9.1 & 9.4\\
$P^{glue}_{away}$ & 0.41 & 0.50 & 0.54 & 0.55\\
\hline
\end{tabular}
\caption{\label{T-Kinbias2} Parameters characterizing the kinematic and parton type bias for a 20-40 GeV trigger range for hadron and jet triggers (see text).}
\end{table*}

One can infer from these numbers a rather complex picture. The kinematic bias is in general stronger for a hadron trigger than for a jet trigger, but this is a feature already present in vacuum. It needs on average a 20.4 GeV parton to produce a hadron in the 10-15 GeV momentum range, but just 18.2 GeV are sufficient for the production of a jet in the same energy range. However, the additional medium-induced shift $\Delta \epsilon_{med}$ is of the same order of magnitude for both the jet and the hadron trigger and about 1.1 GeV.

As expected, the width of the momentum distribution is somewhat narrower for a jet trigger, in other words a jet has a closer correlation with the original momentum than a hadron even for the rather biased jet definition discussed here. However, even the jet-h correlation probes a $\sim 8$ GeV wide distribution of underlying parton energies. 

The parton type bias in vacuum is clearly much stronger for hadron triggered correlations, in good agreement with the expectation that the softer fragmentation pattern of gluons is less of an issue for jets where sufficiently hard subleading hadrons are simply clustered back. There is little evidence for 'gluon filtering', i.e. a strong additional medium-induced parton type bias for hadron triggers, but this is caused by the low probability to trigger on a gluon jet even in vacuum. For jet triggers, the gluon filtering effect in the medium becomes apparent. 

All in all, the medium-induced kinematic bias, i.e. a shift of the average away side parton spectrum upward by about 1 GeV, is not a huge effect when compared to the geometrical bias (which forces a long in-medium path and hence a significant widening and softening of the fragmentation pattern). The kinematic bias alone would lead to a yield enhancement by about 10\% --- this is to be compared with the $\sim$ 80\% suppression of the away side yield due to the medium-modified fragmentation. This is the reason why the net medium effect is a suppression of the high $P_T$ away side yield.

\subsection{The high-tower trigger bias}

The requirement of having at least one tower with 6 GeV or higher in the event is somewhat troublesome for theoretical calculations. For MC modelling, it implies some measure of inefficiency, as a significant fraction of events has to be discarded after they have been simulated down to the hadronic level. For analytical calculations which are unable to obtain an event-by-event representation of jets on the hadron level at all, the condition is impossible to account for. It is therefore of some interest to assess its importance.

An instructive way to illustrate the effect of requiring a high $P_T$ particle in the event is to plot $P(z_{jet})$, the probability of recovering $E_{jet} = E_0 z_{jet} $ given an initial parton with energy $E_{0}$ (with the convention $E_{jet} = 0$ if no jet or no particle above 6 GeV has been found).

\begin{figure*}[htb]
\begin{center}
\epsfig{file=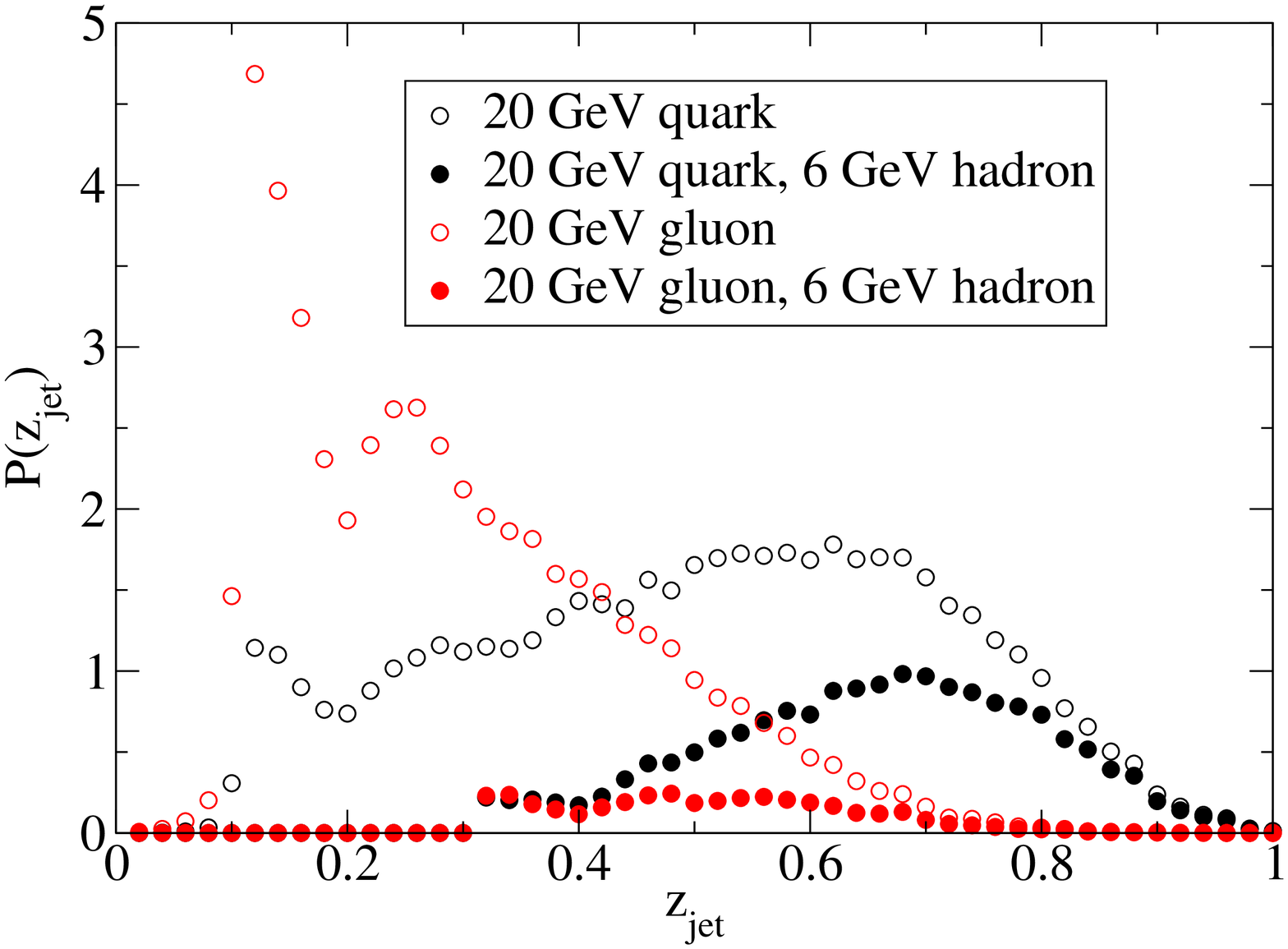, width=7cm}\epsfig{file=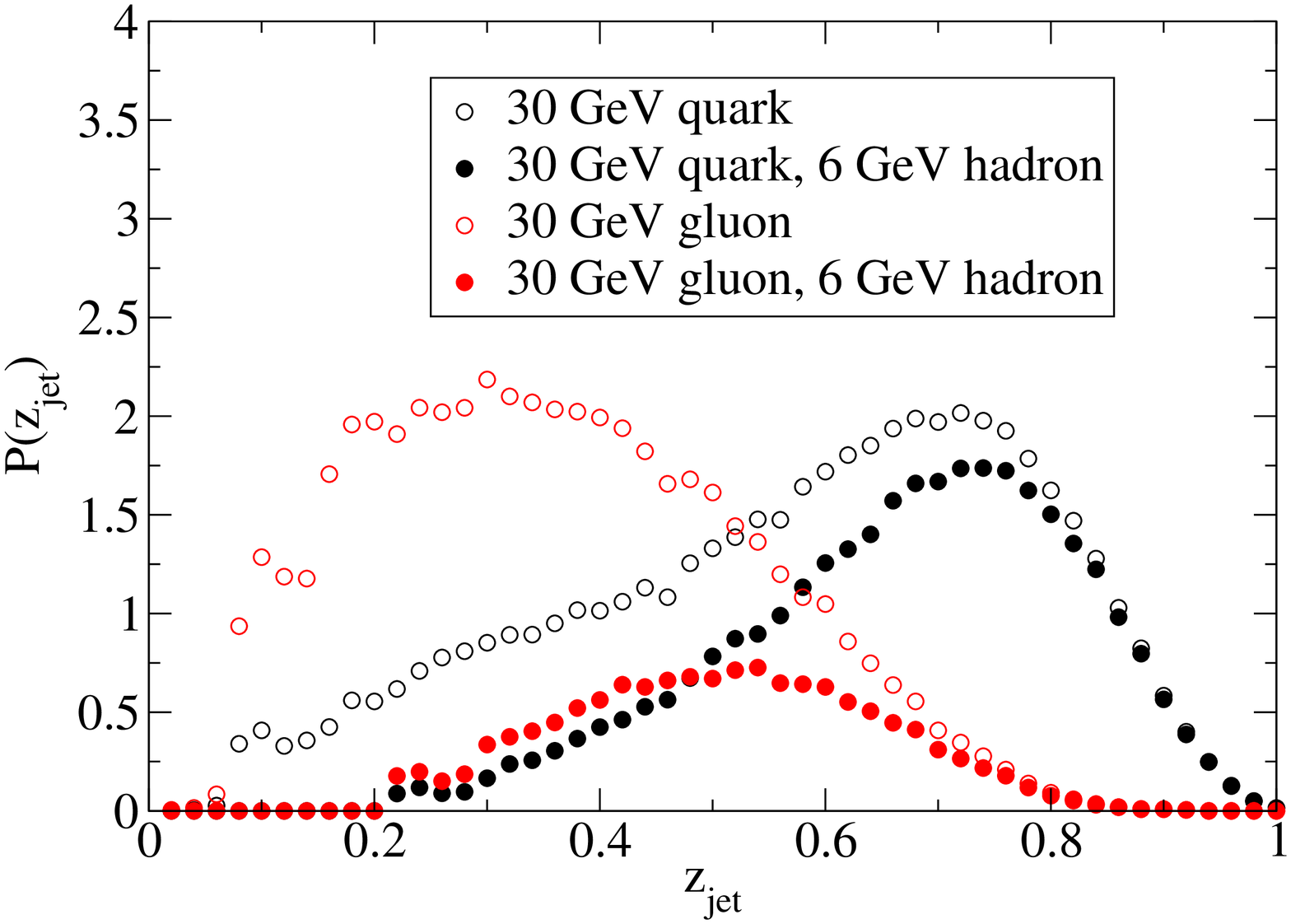, width=7cm}
\end{center}
\caption{\label{F-Trackbias}Probability to recover the fraction $z_{jet}$ of the original parton energy within experimental cuts for 20 GeV partons (left) and 30 GeV partons (right).}
\end{figure*} 

In Fig.~\ref{F-Trackbias} $P(z_{jet})$ for the STAR jet cuts is shown for a 20 GeV and a 30 GeV quark fragmenting in vacuum (where probability-conserving $\delta$-functions at the origin accounting for the case that no particle above 6 GeV has been found and hence no jet is counted are suppressed). The generic trend is that for hard and collimated jets where $z_{jet} \rightarrow 1$ the hard particle requirement does not make any difference, whereas dramatic effects are observed at lower $z_{jet}$, especially in the gluon case where the majority of fragmenting 20 GeV gluons does not lead to any hadron above 6 GeV. As expected, the effects lessen with increased parton energy.

\begin{figure*}[htb]
\begin{center}
\epsfig{file=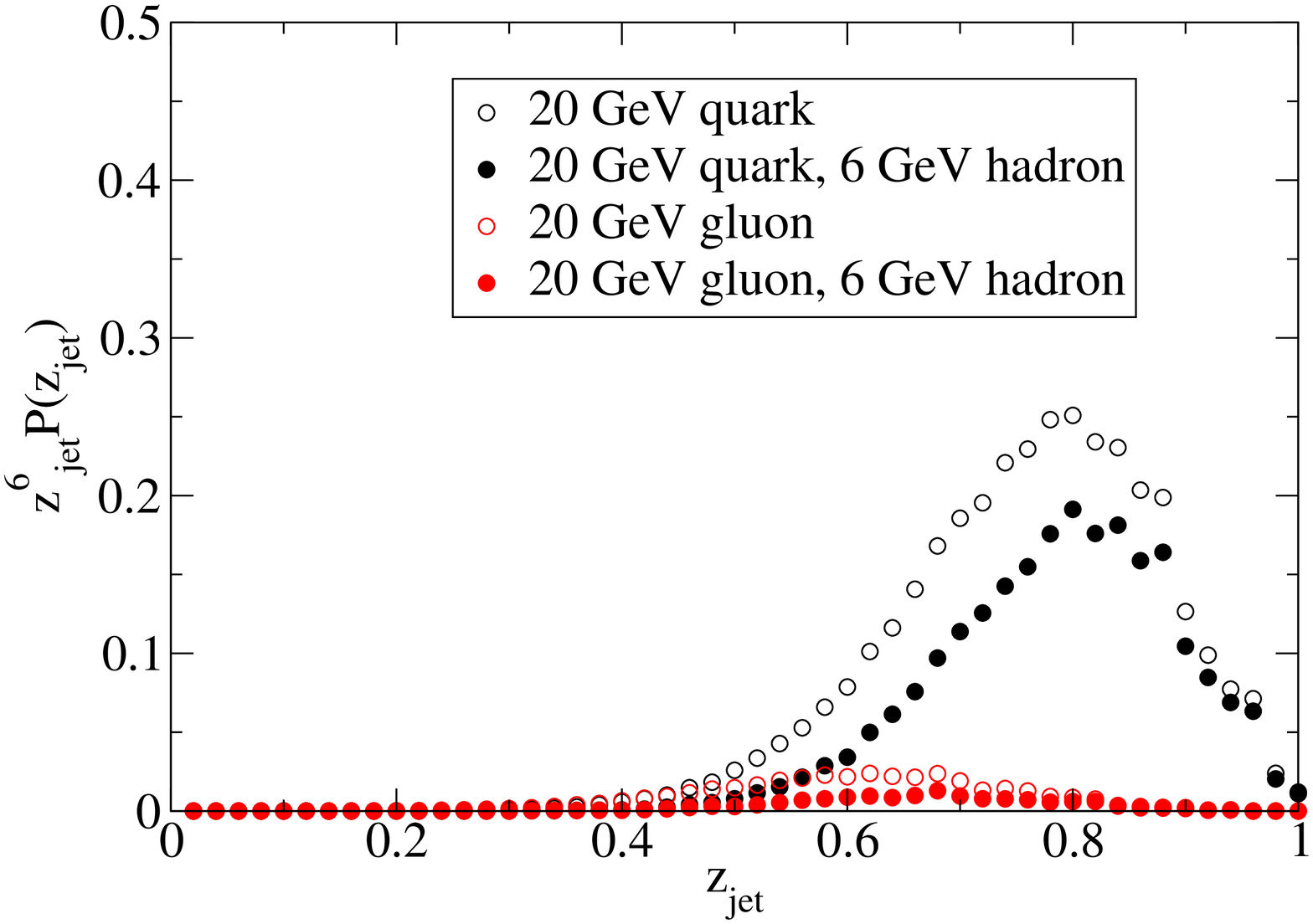, width=7cm}\epsfig{file=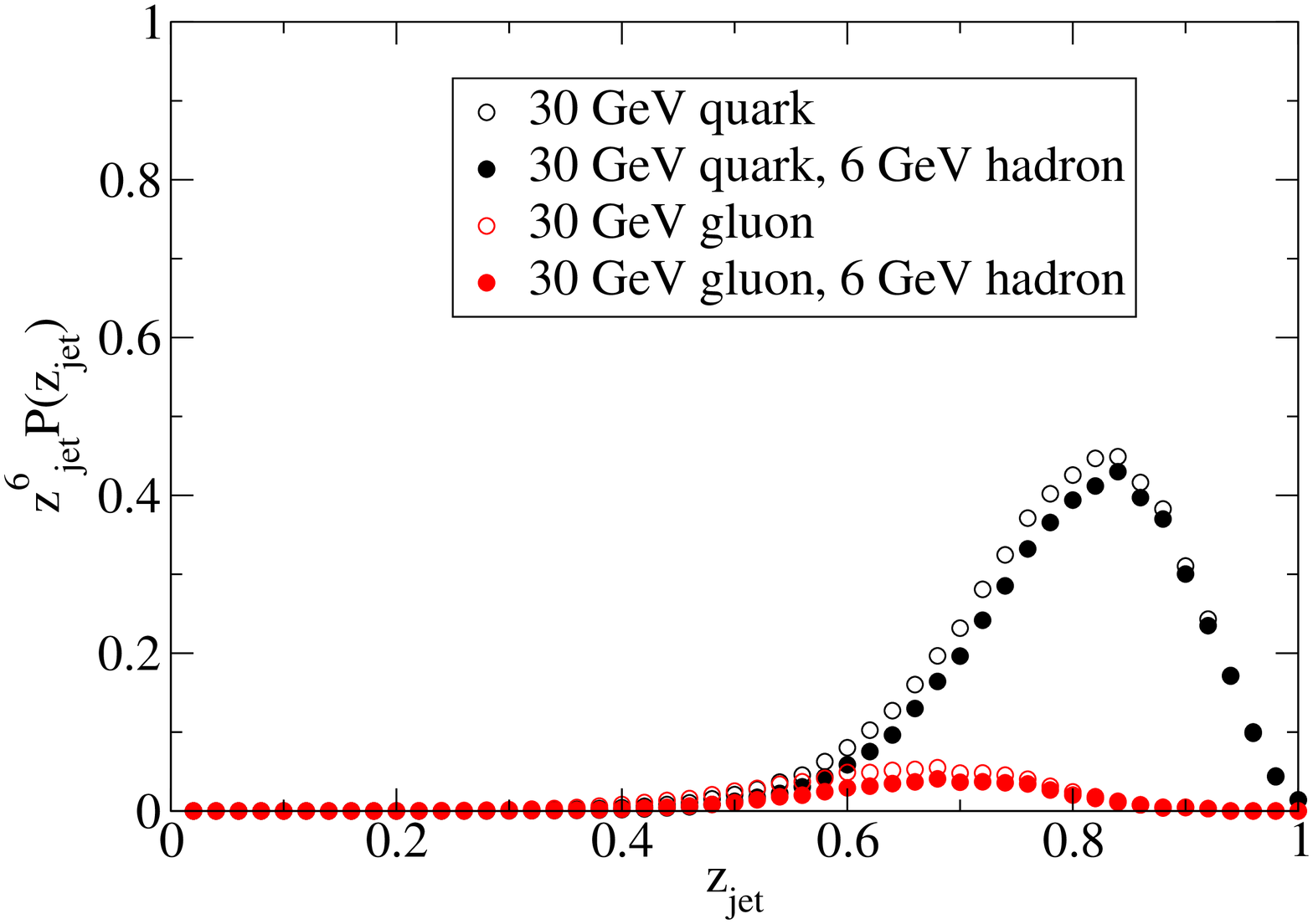, width=7cm}
\end{center}
\caption{\label{F-Trackbias2}$P(z_{jet}|\,20 \,\text{GeV})$ (left) and $P(z_{jet}|\,30\, \text{GeV})$ (right), weighted with $z_{jet}^6$ to demonstrate the effect of sampling the distribution with a steeply falling spectrum. Note that $P(z_{jet}|E_0)$ in reality has a weak scale dependence on $E_0 = E_{jet}/z_{jet}$ and the fact that here the scale is here fixed independent of $z_{jet}$ is approximating the problem for the sake of illustration only.}
\end{figure*}

However, in practice events with low $z_{jet}$ are unlikely to pass the trigger cut, as they require comparatively rare partons with high energy such that $z_{jet}E_0 > E_{trigger}$. Rather, the distribution is weighted by the primary parton spectrum. In order to roughly illustrate the effect of this weight, the quantity $z_{jet}^6 P(z_{jet})$ (which assumes that $P(z_{jet})$ evolves only slowly with $E_0$ and approximates the RHIC parton spectrum by a power law) is shown in Fig.~\ref{F-Trackbias2}.

It is evident that hard and collimated fragmentation where the differences between the presence or absence of a hard particle are small is the favoured situation, which means that the high tower trigger bias is in practice much suppressed by the steeply falling parton spectrum.

In the full MC simulation, the high tower trigger condition translates in the end into an additional kinematic bias of a $\sim 0.5$ GeV average upward shift of the away side parton energy. As discussed above, such a shift is a small effect when compared to the huge suppression caused by the geometry bias. To the accuracy to which the in-medium evolution of parton showers can be currently computed, neglecting the high tower requirement is thus a valid approximation.

\section{Away side jet structure observables}

Let us now focus on the jet structure as imaged via correlations on the away side. The longitudinal momentum distribution is probed in $I_{AA}$ and the momentum balance function $D_{AA}$ whereas the Gaussian width of the correlation is a probe for the transverse jet structure.

\subsection{Longitudinal jet structure}

\begin{figure}[htb]
\begin{center}
\epsfig{file=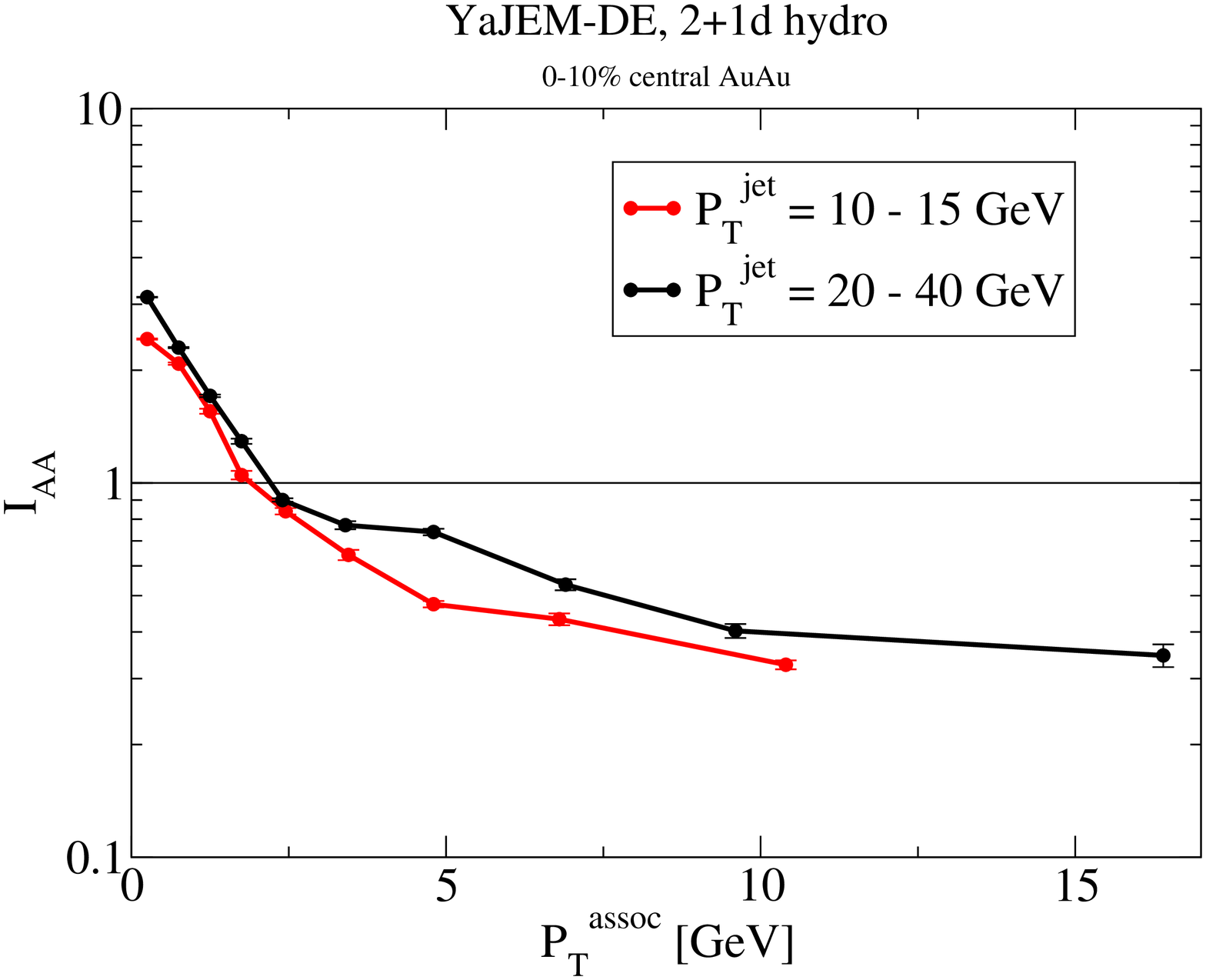, width=7cm}
\end{center}
\caption{\label{F-IAA}Nuclear suppression factor for the conditional away side yield $I_{AA}(P_T)$ computed with YaJEM-DE for 0-10\% central 200 AGeV Au-Au collisions for two different jet trigger energy ranges.}
\end{figure}

The modification factor for the conditional away side yield, $I_{AA}(P_T)$ is shown in Fig.~\ref{F-IAA}. The result shows an enhancement of the yield at low $P_T$ and a suppression of the yield at high $P_T$. This is consistent with the interpretation of energy loss from the leading parton as perturbative production of soft gluons, leading to additional soft hadron yield after hadronization. 

A similar pattern has been observed in the comparison of YaJEM with h-h correlations \cite{YaJEM-DE,h-h}. Given the similarity of the kinematic bias, this should not come as a surprise. Based on the results of \cite{YaJEM-DE}, the quantity can be expected to be sensitive to the amount of direct energy loss into the medium via $\hat{e}$ in a similar way as $I_{AA}(z_T)$ in h-h correlations, although we do not demonstrate this again in this work.

A rather remarkable observation is that the crossing point of $I_{AA}$ with unity is quite independent of the trigger energy range (and by the arguments given in the previous section, hence independent of the away side parton energy). This indicates that the medium-modified fragmentation function (MMFF) is not modified at a constant momentum fraction $z = E_{had}/E_0$ but rather at a constant energy scale. In other words, the medium modification can manifestly not be cast into a modified probabilistic branching kernel $P'(z)$ as has been assumed in several models \cite{BW,Q-PYTHIA}. The same observation has now also been made at LHC for jets at 100 GeV where a pronounced enhancement of the jet fragmentation function below 3 GeV was observed \cite{CMSFF}.

Such a modification of jets confined to a low $P_T$ range independent of jet energy  has been observed in YaJEM already in \cite{JetShape}. The relevant difference in modelling is that \cite{BW,Q-PYTHIA} try to simplify the problem by casting all medium modification into a modification of splitting probabilities of virtual partons while energy and momentum \emph{inside the jet} remains exactly conserved whereas YaJEM assumes an explicit exchange of energy and momentum between jet and medium and does not require momentum conservation inside the jet alone. This allows the typical momentum scale of the medium to appear explicitly in the modelling: The jet structure changes dramatically as soon as the momentum cumulatively transferred from the medium is of the order of a shower parton momentum and allows for a significant deflection. Based on this argument, we may expect a modification of the transverse jet structure at the same scale of $\sim 3$ GeV.

The high $P_T^{assoc}$ behaviour of $I_{AA}(P_T)$ shows an almost momentum-independent suppression. This behaviour can be understood by the same argument as above: In YaJEM, the parton splitting kernels $P_{i\rightarrow j,k}(z)$ who iteratively generate the shower and hence the fragmentation function are not directly modified. This implies that in the regime where the momentum transfer from the medium can not significantly alter parton kinematics, the fragmentation function necessarily must be self-similar with the same shape as observed in vacuum \cite{HPproc}, and thus energy loss from leading partons can only result in an apparent suppression of the longitudinal momentum distribution, but not acquire any $P_T$ dependence. 

\begin{figure}[htb]
\begin{center}
\epsfig{file=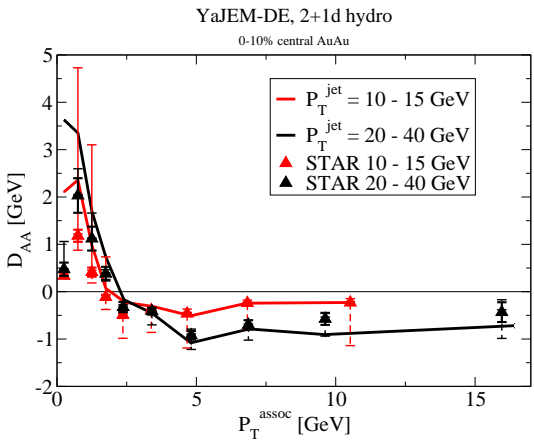, width=7cm}
\end{center}
\caption{\label{F-DAA}Away side momentum balance function $D_{AA}(P_T)$ computed with YaJEM-DE for 0-10\% central 200 AGeV Au-Au collisions for two different jet trigger energy ranges, compared with preliminary STAR data\cite{STAR-jet-h}.}
\end{figure}

In Fig.~\ref{F-DAA} the momentum balance function $D_{AA}(P_T)$ is shown in comparison with preliminary STAR data \cite{STAR-jet-h}. Good agreement with the data within statistical and systematical errors is observed for all $P_T$ but the lowest bin. This again reflects the dynamics of momentum lost from partons at high $P_T$ predominantly appearing in low $P_T$  additional hadron production.

Both observables reflect the medium-modified fragmentation function of jets where the parent parton momentum is averaged over the distribution in Fig.~\ref{F-Kinbias} and the path through the medium taken by the away side parton is averaged over the geometry shown in Fig.~\ref{F-Geobias}, i.e. a rather strong modification.

\subsection{Transverse jet structure}

\begin{figure*}[htb]
\begin{center}
\epsfig{file=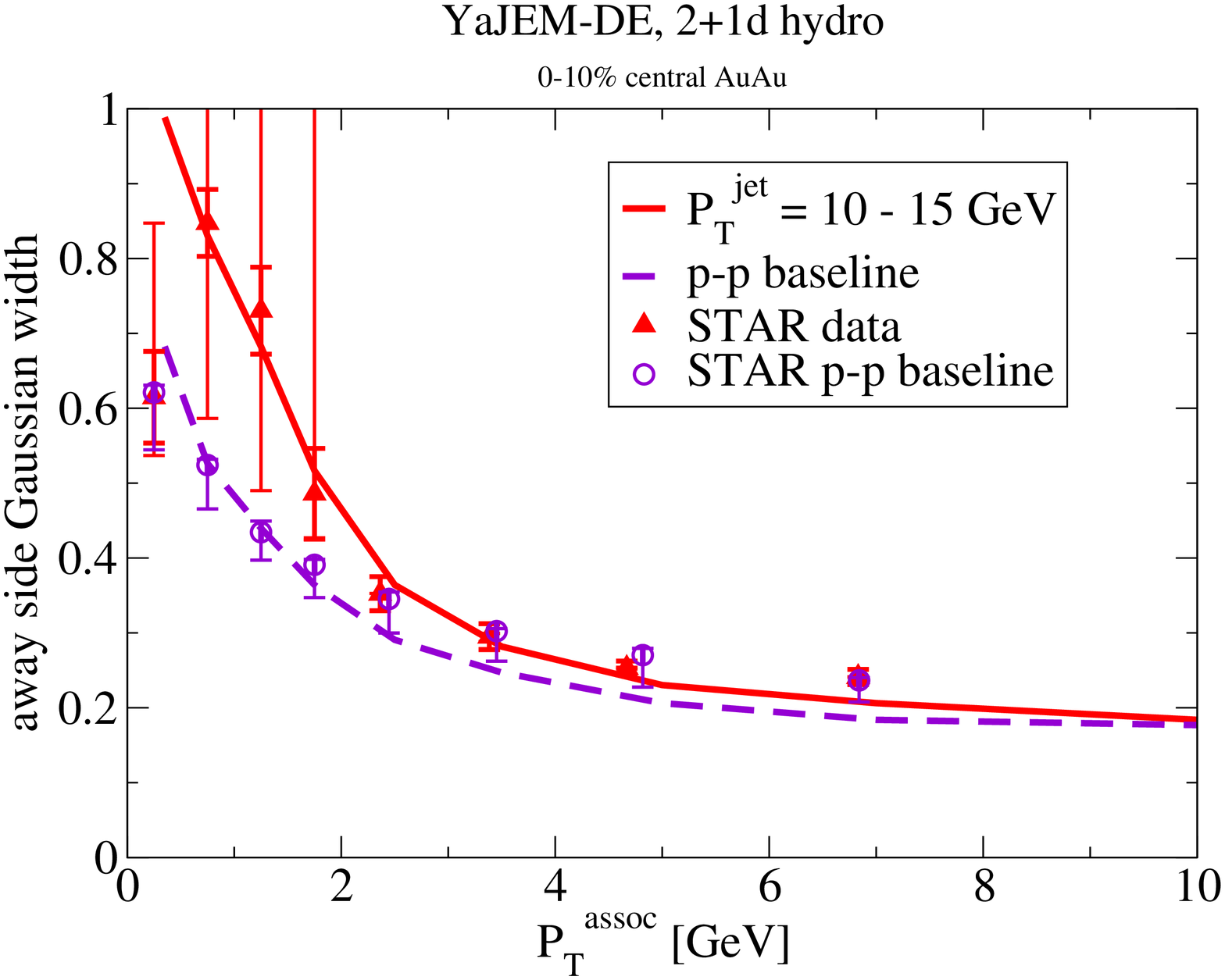, width=7cm}\epsfig{file=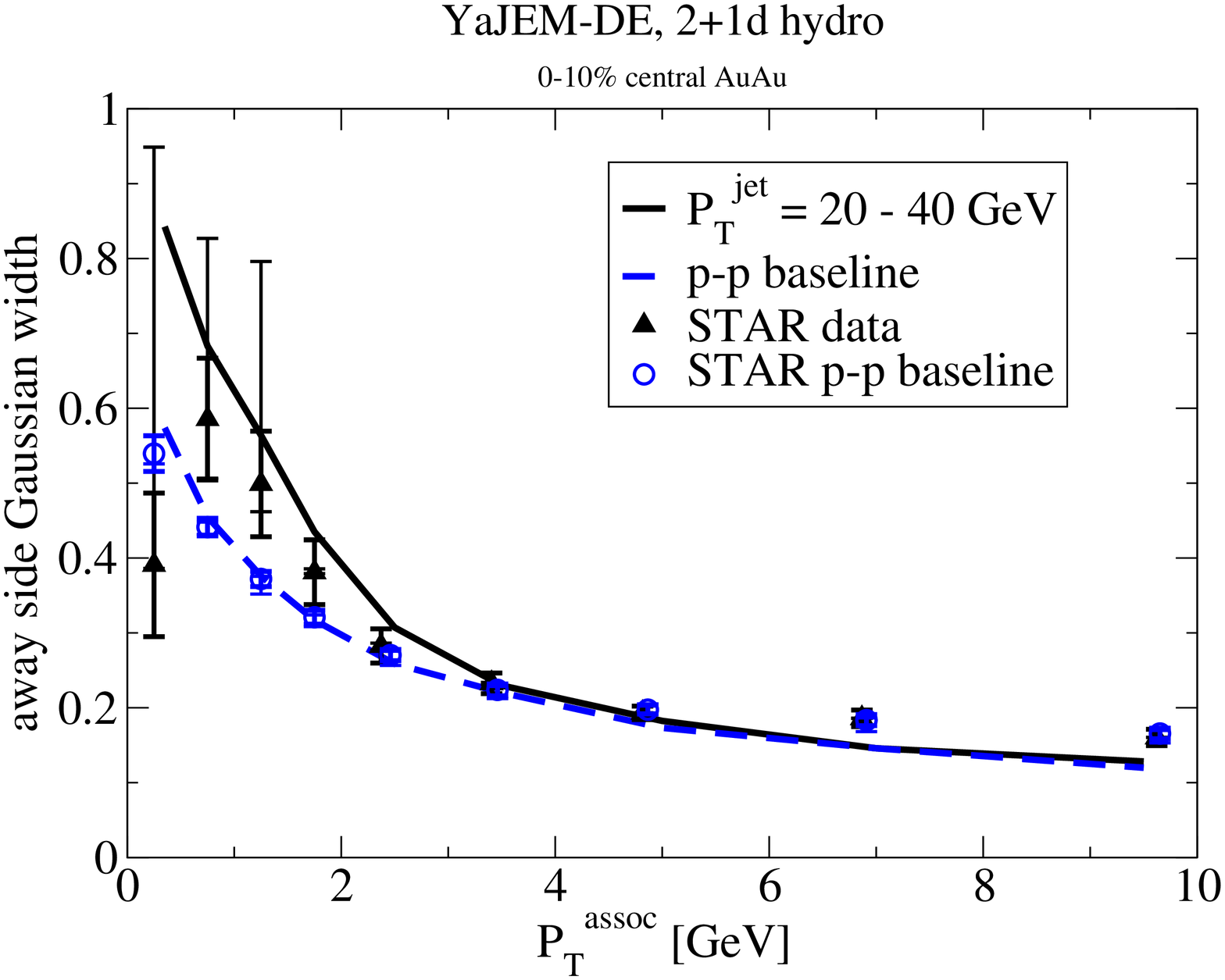, width=7cm}
\end{center}
\caption{\label{F-Gaussian}Gaussian angular width of the away side correlation peak as a function of $P_T$ as computed with YaJEM-DE for 0-10\% central 200 AGeV Au-Au collisions for two different jet trigger energy ranges (left and right panel), compared with the p-p baseline width for the same energy without medium modification and preliminary STAR data \cite{STAR-jet-h}}
\end{figure*}

In Fig.~\ref{F-Gaussian}, the Gaussian angular width of the away side correlation as a function of $P_T$ is shown in comparison with the preliminary STAR data \cite{STAR-jet-h}. The width of the away side correlation structure arises as a combination of two distinct effects: 1) the imbalance in momentum on the level of the shower-initiating parent partons ($k_T$-broadening) and 2) the spread of individual hadrons inside the jet around the axis defined by the parent parton ($j_T$-broadening). Given that $k_T/E_{0}$ is not a large quantity, $k_T$-broadening is only dominant al high $P_T^{assoc}$ where the shower is highly collimated.

Good agreement within statistical and systematic errors throughout the whole $P_T$ range is obtained, with some deviations of the baseline calculation at high $P_T$. This presumably implies that the value of $k_T$ is even larger than assumed here.

 It can be seen in the figure that at high $P_T^{assoc}$ down to about 3 GeV the vacuum and the medium-modified result are fairly similar, but dramatic differences in width appear at lower momenta. The momentum scale of the changed behaviour is consistent with the scale at which the longitudinal modification of the jet structure changes its behaviour from suppression to enhancement. This is in agreement with the idea outlined above that intra-jet momentum can only significantly been altered at a momentum scale set by the medium.

The combination of $D_{AA}(P_T)$ and the Gaussian width is a very differential characterization of the longitudinal and transverse momentum distribution inside medium-modified jets even down to low $P_T$ which is difficult to access directly with reconstructed jets. The characterization of the transverse distribution in terms of the Gaussian width is in fact superior to observables such as the jet shape which is dominated by the high $P_T$ dynamics and tends to hide the fact that the most dramatic modifiactions occur at low $P_T$.

A further characterization of the jet structure would need to probe for intra-jet correlations such as the subjet structure. This could in principle be done with jet-triggered dihadron correlations, however we will not pursue this idea here further.

\section{Further aspects of trigger conditions}

Let us now turn to a closer investigation of the potential of jet triggers which can be exploited by varying the trigger conditions or the kinematic conditions. 

\subsection{The geometry bias}

\begin{figure}[!htb]
\begin{center}
\epsfig{file=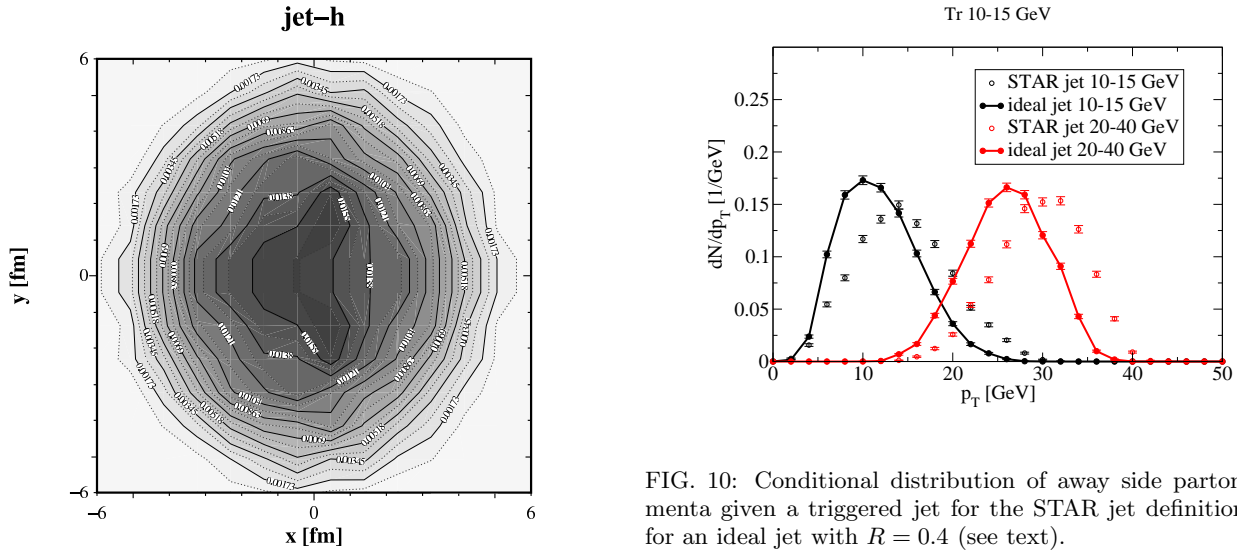, width=7cm}
\end{center}
\caption{\label{F-Geobias1}Comparison of the probability density of a vertex in the transverse $(x,y)$ plane to fulfill a 10-15 GeV trigger condition in 0-10\% central 200 AGeV Au-Au collisions for an ideal jet trigger (see text). The trigger parton moves into the $-x$ direction.}
\end{figure}

One of the fascinating aspects of using a jet triggered correlation is that the jet definition allows to dial the amount of geometrical bias. In Fig.~\ref{F-Geobias}, a very strong geometrical bias is observed. In contrast, in Fig.~\ref{F-Geobias1} the geometrical bias is shown for the same calculation with the only difference that the jet definition is changed to an jet definition for which all particles at all $P_T$ are clustered with anti-$k_T$ with $R=0.4$ (i.e. no PID cuts, no $P_T$ cut). 

The result is an almost unbiased distribution of vertices with $s=1.13$. While this situation is experimentally not accessible at RHIC, similarly unbiased jet definitions can easily be used by the LHC experiments. Thus, by increasing the particle $P_T$ cuts, a very weakly surface biased situation can be turned into a highly surface-biased situation, which can be used to dial the expected amount of medium modification on the away side. 

\begin{figure}[htb]
\begin{center}
\epsfig{file=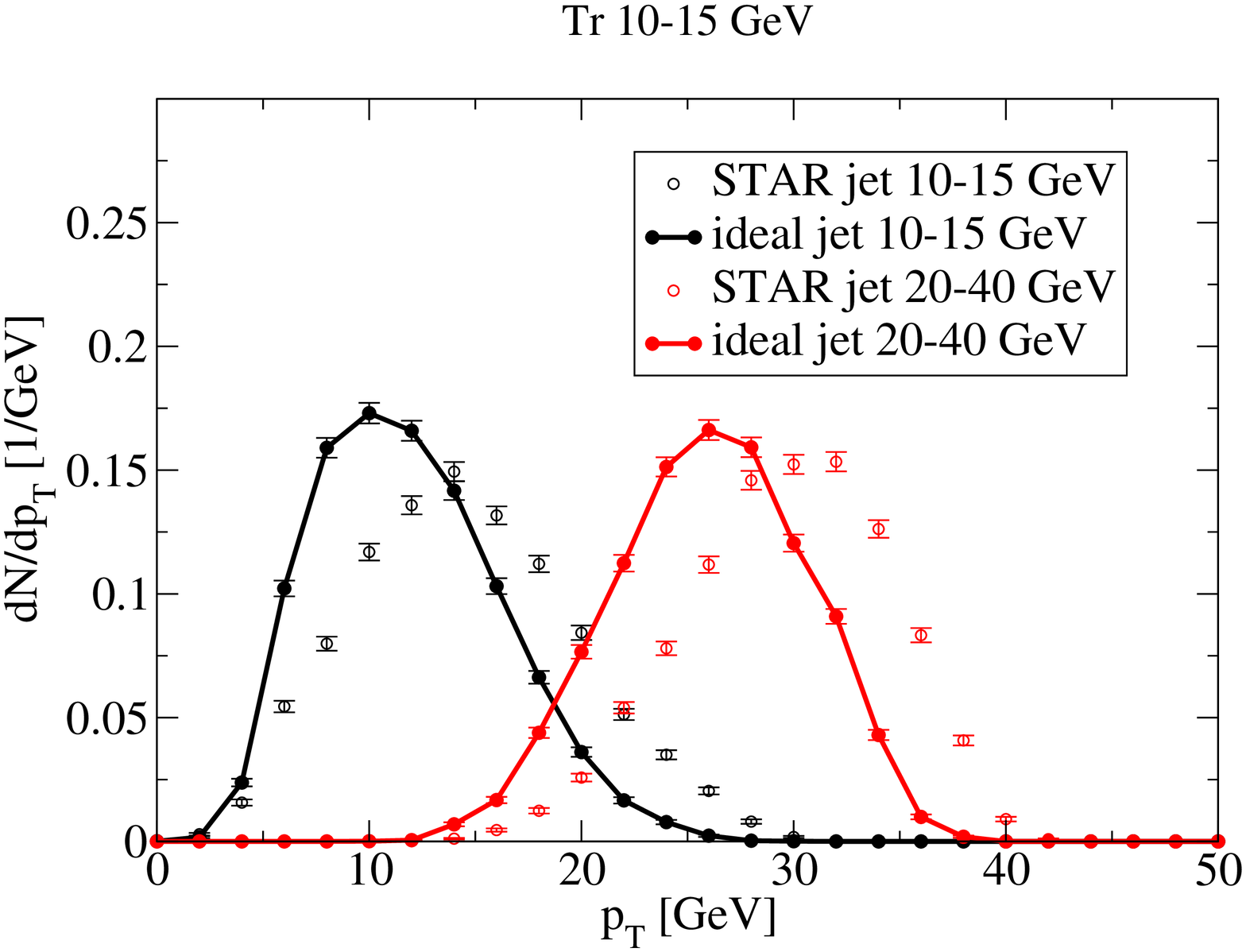, width=7cm}
\end{center}
\caption{\label{F-Kinbias1}Conditional distribution of away side parton momenta given a triggered jet for the STAR jet definition and for an ideal jet with $R=0.4$ (see text).}
\end{figure} 

However, it is important to realize that geometry bias and kinematic bias cannot be varied independently --- the higher energy fraction recovered by the ideal jet as compared with the STAR jet definition has implications for the conditional away side parton yield as well, which is shown in Fig.~\ref{F-Kinbias1}. In essence, using a jet definition which captures more of the original parton energy on the near side implies a downward shift of the mean momentum of the away side partons by several GeV --- since on average less energy is needed to fulfill the trigger condition on the near side, the recoiling parton also will have less energy. It is important to understand the interplay of these effects properly before making a comparison between triggers using different jet definitions, even if the same trigger momentum range is used.

\subsection{The role of the pQCD parton spectrum}

A cornerstone of several arguments presented above was the fact that for a steeply falling parton spectrum fragmentation is strongly forced to be hard and collinear by imposing a trigger condition, as the situation that a rare hard parton undergoes soft fragmentation is very suppressed. One of the consequences is a relatively good correlation between trigger momentum range and actual away side parton energy distribution.

However, when going to higher $\sqrt{s}$ where the spectral shape flattens, this argument applies increasingly less. In order to illustrate the importance of this effect in isolation, we compare the simulation for RHIC conditions with a situation in which only the parton spectrum is computed for LHC conditions, everything else is kept fixed (in reality, also intrinsic $k_T$ and most important the medium density is expected to change).

\begin{figure}[htb]
\begin{center}
\epsfig{file=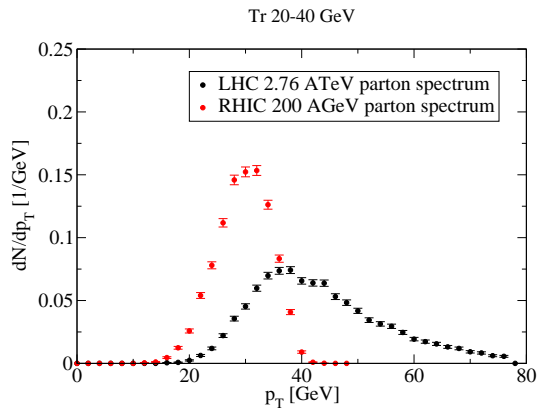, width=7cm}
\end{center}
\caption{\label{F-Kinbias2}Conditional distribution of away side parton momenta given a triggered jet assuming the kinematical conditions at RHIC vs. LHC.}
\end{figure} 

One can easily see that the qualitative argument given above is correct --- the correlation between trigger momentum range and away side parton momentum weakens significantly, and a long tail of high $P_T$ partons contributes to the away side yield, complicating the interpretation of any away side measurement which represents then an average over a wide momentum range. From this perspective, the steeply falling parton spectrum at RHIC constitutes actually an advantage over LHC kinematic conditions.

\section{Conclusions}

On the conceptual side, jet-h correlations offer a number of advantages. The use of a jet trigger as compared to a hadron or even $\gamma$ trigger allows experiments to collect much higher statistics since the rate of jets into a given $P_T$ range is higher than the rate of hadrons or photons, and this in turn allows differential studies of the away side. At least for RHIC kinematics, there is a reasonably good correlation between jet trigger energy range and the underlying parton energy range which is probed, however this is no longer the case at LHC --- here presumably $\gamma$-h correlations are needed to constrain parton kinematics.

At the same time, jet triggers appear very versatile tools which can be engineered to lead to a certain geometrical bias by a suitable choice of the jet constituent $P_T$ cut. In simulations, both an almost unbiased distribution and a distribution biased beyond what is seen for hadron triggered events could be achieved. 

Measuring the correlation of hadrons on the away side allows to probe the longitudinal and transverse single particle distributions of jet constituents down to very low $P_T$ and out to large angles, which is a particular advantage for tracing the medium-induced modification to jet structure. In this, a correlation measurement is superior to jet finding on the away side, as jet finding in an A-A environment is limited in its ability to reach to large angles and low $P_T$. In principle, in order to access the medium-modification of intra-jet correlations and to probe physics like a modified subjet structure or modifications of angular ordering \cite{AntiAngularOrdering}, correlations of a trigger with two away side particles can be used.

On the physics side, the longitudinal and transverse jet structure of modified jets as measured by $D_{AA}(P_T)$ and the angular Gaussian width is well described by YaJEM-DE except in the very low $P_T$ region where the physics is not dominated by pQCD and the model is expected to fail. Thus, the observed jet modification is well in line with the general idea that the medium opens additional kinematical phase space for radiation, the induced soft radiation is rapidly decorrelated by subsequent interactions with the medium while a small part of the energy lost from hard partons directly excites medium degrees of freedom. The combination of these mechanisms leads to apparently unmodified but rate-suppressed jets above a scale of $\sim 3$ GeV and a wide-angle, soft plateau-like structure below this scale.

Of particular interest for determining the precise nature of the interaction of hard partons with the bulk medium is the origin of the scale $P_{med} \approx 3$ GeV. It is certainly consistent with a back-of-the-envelope estimate that the scale is given by the typical accumulated medium momentum probed during subsequent interactions $P_{med} = L/\lambda \langle P \rangle$. Choosing a typical length $L=5$ fm, a mean-free path $\lambda = 1$ fm and for the typical momentum scale in the medium $\langle P \rangle = 3T$ with the medium temperature $T = 200$ MeV leads to $P_{med} \approx 3$ GeV. However, in this case it would be very interesting to demonstrate the change of the scale by experimentally varying temperature (e.g. by comparing RHIC and LHC) or by varying mean free path. An alternative position is that $P_{med}$ is set by strong coupling physics not accessible via pQCD arguments. Future reaction plane differential measurements of jet-h correlations at RHIC and LHC might be a suitable way to distinguish these scenarios and to establish in detail what aspects of jet physics are governed by pQCD and what aspects by strong coupling QCD.

\begin{acknowledgments}
 
I would like to thank Alice Ohlson, Helen Caines, Megan Connors and J\"{o}rn Putschke for long discussions and email exchanges leading to this paper. This work is supported by the Academy researcher program of the
Academy of Finland, Project No. 130472. 
 
\end{acknowledgments}

\end{document}